\documentclass[aps,preprint,showpacs,floatfix]{revtex4}
\usepackage{graphicx,bm,amsmath,dcolumn}
\graphicspath{{figures}}
\begin{document}
\newcommand{\half}{\frac{1}{2}}
\newcommand{\ith}{^{(i)}}
\newcommand{\im}{^{(i-1)}}
\newcommand{\gae}
{\,\hbox{\lower0.5ex\hbox{$\sim$}\llap{\raise0.5ex\hbox{$>$}}}\,}
\newcommand{\lae}
{\,\hbox{\lower0.5ex\hbox{$\sim$}\llap{\raise0.5ex\hbox{$<$}}}\,}
\newcommand{\be}{\begin{equation}}
\newcommand{\ee}{\end{equation}}
\newcommand{\bea}{\begin{eqnarray}}
\newcommand{\eea}{\end{eqnarray}}

\title{Critical frontier for the Potts and percolation models on triangular-type and kagome-type lattices II: Numerical analysis}
\author{Chengxiang Ding$^{1}$, 
Zhe Fu$^{1}$, 
Wenan Guo$^{1}$, 
and F. Y. Wu$^{2}$} 
\affiliation{$^{1}$Physics Department, Beijing Normal University,
Beijing 100875, P. R. China }
\affiliation{$^{2}$ Department of Physics, Northeastern University, Boston, Massachusetts 02115, USA }
\date{\today} 
\begin{abstract}
In a recent paper (arXiv:cond-mat/0911.2514), one of us (FYW) considered the Potts model and bond and site percolation on two general classes of two-dimensional
lattices, the triangular-type and kagome-type lattices, and obtained closed-form expressions for the critical
frontier with applications to various lattice models. For the triangular-type lattices
Wu's result is exact, and for the kagome-type lattices Wu's  expression is  under a 
homogeneity assumption. The purpose of the present paper is
two-fold: First, an essential step in Wu's analysis is the derivation of 
lattice-dependent constants $A, B, C$ for various lattice models, a process which can be tedious.  We present here 
a derivation of these constants for  subnet networks  using a computer algorithm. 
Secondly, by means of a finite-size scaling analysis based on numerical  transfer matrix calculations, we 
deduce critical properties and critical thresholds of various models and
assess the accuracy of the homogeneity assumption. Specifically, we analyze
the $q$-state Potts model and the bond percolation on the 3-12 and kagome-type subnet lattices 
$(n\times n):(n\times n)$, $n\leq 4$, for which the exact solution is not known. 
To calibrate the accuracy of the finite-size procedure, we apply the same numerical 
analysis to models for which the exact critical frontiers are  known.
The comparison of numerical and exact
results shows that our numerical determination of critical thresholds is accurate to 7 or 8 
significant digits.
This in turn infers that the homogeneity assumption determines critical frontiers 
with an accuracy of 5 decimal places or higher. Finally, we also obtained the exact percolation
thresholds for site percolation on kagome-type subnet lattices $(1\times 1):(n\times n)$ for $1\leq n \leq 6$.
 \end{abstract}
\maketitle 
\section{Introduction}
The $q$-state Potts model \cite{potts,wfypotts} is a very important model in the study of 
phase transitions and critical phenomena.  
The critical frontier, or the loci of critical points, of the Potts model was first determined by Potts
\cite{potts} for the square lattice.
The critical exponents of the Potts model are obtained by conjectures on the basis of numerical
evidence and by using Coulomb gas theory \cite{dNijs, BE, NRS, NBRS, dNijs1}. 
According to the universality  hypothesis \cite{griffiths}, the Potts model on  different lattices belongs
to the same universality class.
But the determination of the critical frontier of the $q$-state Potts model
in general, which includes the $q=1$ bond and site percolation, is still an outstanding challenge.  
Particularly, the threshold of site percolation has remained largely unresolved.

\begin{figure}[htpb]
\includegraphics[scale=0.4]{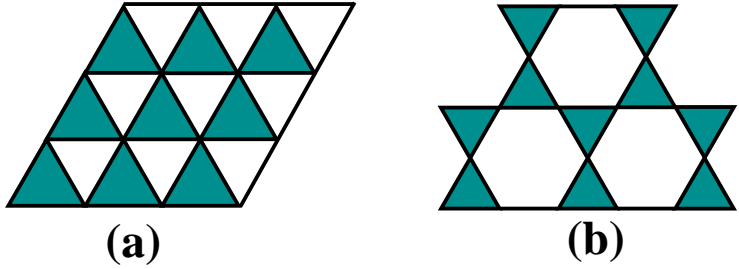}
\caption{(a) Triangular-type lattice. (b) Kagome-type lattice.}
\label{tri&kag}
\end{figure}

\begin{figure}[htpb]
\includegraphics[scale=0.4]{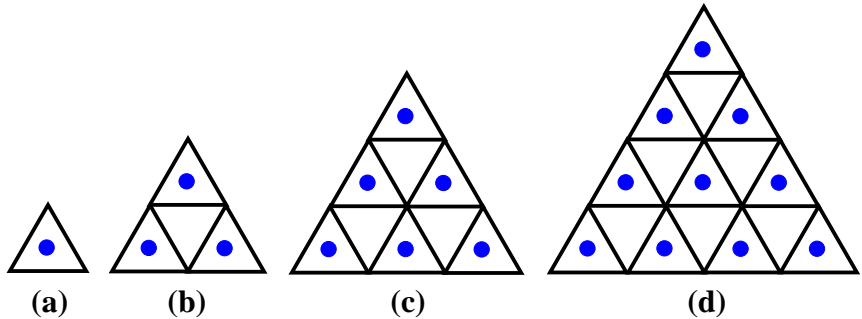}
\caption{Subnet networks. (a) $1\times 1$ subnet. (b) $2\times 2$ subnet. (c) $3\times 3$ subnet. 
(d) $4\times 4$ subnet. The dots denote triangles with 3-site interactions.}
\label{subnets}
\end{figure}

In a recent paper \cite{paperI}, hereafter referred to as I, one of us (FYW) considered the Potts model on 
two classes of very general two-dimensional lattices, the triangular-type and kagome-type lattices shown in 
Fig. \ref{tri&kag}. The Boltzmann weights $W$ of the hatched triangles 
denote interactions involving 3 spins $\tau_1,\tau_2,\tau_3=1,2,\dots,q$ surrounding a triangle,
and are given by
\begin{eqnarray}
W_{\bigtriangleup} (1,2,3)  &=& A+B(\delta_{12}+\delta_{23}+\delta_{31})+C\delta_{123} \, ,\nonumber \\
W_\bigtriangledown (1,2,3) &=& A' + B'(\delta_{12}+\delta_{23}+\delta_{31})+C'\delta_{123}\, , 
\label{ABCdef}
\end{eqnarray} 
where $\delta_{ij} = \delta_{\rm Kr}(s_i, s_j), \delta_{123}=\delta_{12}\,\delta_{23}\,\delta_{31}$,
and $A,B,C, A',B',C'$ are constants. 
Spin interactions within the hatched areas can be either 2- or 3-site couplings. 
The hatched triangles can have internal structures 
such as the stack-of-triangle subnets, which are of recent interest \cite{trianglesubnets,kagomesubnets}, 
shown in Fig. \ref{subnets}.
We refer to these structures as {\it subnet networks}. These stack-of-triangle lattices are called {\it subnet lattices}. 
Examples of triangular subnet lattices and kagome-type subnet lattices are shown in 
Fig. 2 and 3 of I.
The $1 \times 1$ subnet lattices are the triangular and kagome lattices themselves. We
shall call a kagome-type lattice with $m \times  m$ down-pointing and $n \times n$ up-pointing subnets
an $(m \times m) : (n \times n)$ subnet lattice. 

In I, Wu derived closed-form expressions for the critical frontier of the $q$-state Potts model
for the 2 types of lattices in Fig. \ref{tri&kag}. For the triangular-type lattices the critical 
frontier is exact, but for the kagome-type lattices the critical frontier is obtained under a
homogeneity assumption. 

The purpose of this paper is two-fold:

First, an essential step in Wu's analysis is the derivation of relevant lattice-dependent 
constants $A, B, C$ for subnet networks. The derivation, while elementary, is tedious. Here we use
a computer algorithm to evaluate them. Details of the algorithm are described in Sec. \ref{abc}.

Secondly, we determine the critical frontier numerically and
examine the accuracy of the homogeneity assumption.
Specifically, we carry out a finite-size scaling analysis based on transfer matrix calculations
to numerically determine the critical frontier for several lattice models, including the Potts model
on the 3-12 and kagome-type \ $(n\times n):(n\times n)$ \ subnet lattices, 
for which the exact thresholds are not known. To assess the accuracy of the numerical determination, we
also apply the procedure to models
  for which the exact critical thresholds are known. These include
the Ising model and site percolation on the 3-12 lattice and kagome-type \ $(1\times 1):(n\times n)$ \ subnet lattices,
$n\leq 6$.  Comparison of numerical and known exact results 
shows that our numerical procedure is accurate to 7 or 8 decimal places.
This in turn infers that the critical frontier determined using  the homogeneity assumption \cite{paperI}
of I is accurate to 5 decimal places or higher.

Our paper is organized as follows: The main findings of I are summarized in Sec. \ref{frontiers}. 
We describe in Sec. \ref{abc} the algorithm we use to obtain the expressions of $A,B,C$ for the Potts model
with pure 2- and/or  3-site subnet interactions. The
resulting expressions of $A, B,  C$ are listed in the Appendix.
In Sec. \ref{TM}, we describe the transfer matrix technique and the finite-size scaling method. 
Numerical results of our transfer matrix calculations and finite-size scaling analysis are given 
in Sec.  \ref{list_result}. 
New exact thresholds are also given in Sec.  \ref{list_result} for site percolation
on kagome-type $(1\times 1):(n\times n)$ subnet lattices for $n$ up to 6.
We summarize our main findings in Sec. \ref{summary}.

\section{Main results of I}
\label{frontiers}
We summarize in this section the main results of I.
 
For the triangular-type lattice shown in Fig. \ref{tri&kag}(a), the partition function is
\be
Z_{\rm tri} (q; A,B,C) = \sum_{\tau_i=1}^q \prod_\bigtriangleup W_\bigtriangleup (i,j,k)\, , \label{triboltzmann} \\
\ee
where the products are taken over the up-pointing triangles.  
Wu \cite{paperI,qAC} showed that, in the regime
\be
2B+C >0,  \quad 3B+C>0, \label{condition}
\ee
in which the ground state of $W_\bigtriangleup$ is ferromagnetic,
 the exact critical frontier is given by 
\begin{equation}
qA=C \label{CA}.
\end{equation}

\begin{figure}[htbp]
\includegraphics[scale=0.3]{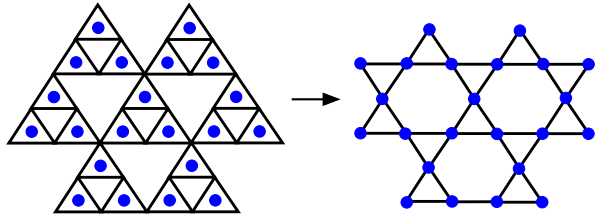}
\caption{Site percolation on the kagome lattice.}
\label{Spkag}
\end{figure}

\begin{figure}[htbp]
\includegraphics[scale=0.25]{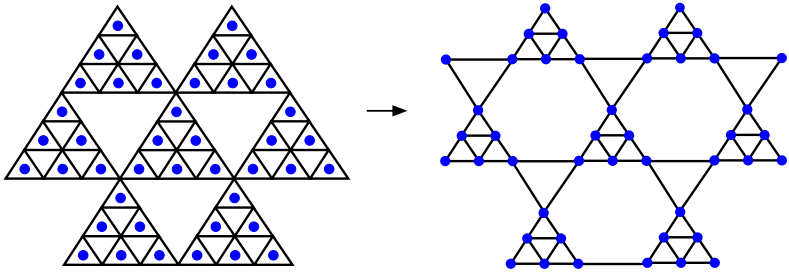}
\caption{ Site percolation on the $(1\times 1):(2\times 2)$ kagome-type subnet lattice.}
\label{akagsubnets}
\end{figure}

The critical function (\ref{CA})  yields the exact thresholds of {\it site} percolation on 
lattices generated from triangular-type lattices. 
Consider a Potts model on a triangular subnet lattice with
pure 3-site interactions in doted triangles shown in Fig. \ref{subnets}.  
Regarding faces of 3-spin interactions in an $n\times n$ subnet as sites of a new lattice, the
Potts model maps to a site
percolation on a \ $(1\times 1):(n-1)\times (n-1)$ \ kagome-type subnet lattice. The critical 
frontier  (\ref{CA}) then gives the exact threshold of the site percolation.
Examples of the mapping are shown in Fig. \ref{Spkag} and \ref{akagsubnets} for $n=2$ and 3.

For kagome-type lattices shown in Fig. \ref{tri&kag}(b) the partition function is
\bea
Z_{\rm kag} (q; A,B,C; A', B', C') &=& \sum_{\tau_i=1}^q \big[\prod_\bigtriangleup W_\bigtriangleup (i,j,k)
\big] \cdot \big[\prod_\bigtriangledown W_\bigtriangledown (i', j', k') \big]. \label{kagomeboltzmann}
\eea
Wu \cite{paperI} obtained its critical frontier 
 \begin{equation}
\label{conj}
(q^2A+3qB+C)(q^2A^{\prime}+3qB^{\prime}+C^{\prime})-3(qB+C)(qB^{\prime}+C^{\prime})-(q-2)CC^{\prime}=0\, ,
\end{equation}
under a homogeneity assumption. 

\begin{figure}[htbp]
\includegraphics[scale=0.3]{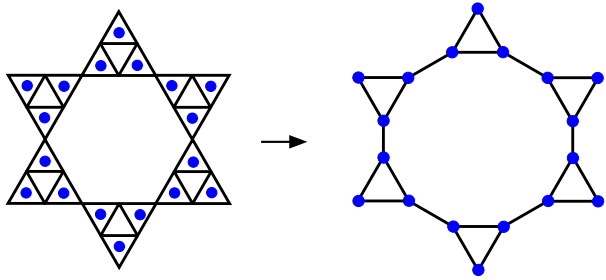}
\caption{Site percolation on the 3-12 lattice.}
\label{sp312}
\end{figure}

The critical point in the case of $q=2$  computed from (\ref{conj}) is exact.
Wu \cite{paperI} also used (\ref{conj}) to compute
  Potts thresholds for the 3-12  and  the \ $(m\times m):(n\times n)$ \ kagome-type subnet 
lattices for $m,n \leq 4$ for which the exact thresholds are not known. 
In addition, Wu   deduced the known exact threshold of site percolation on
the 3-12 lattice by considering the Potts model on the \ $(2\times 2):(2\times 2)$ \ kagome-type lattice 
 as shown in Fig. \ref{sp312}. 
In this case the homogeneity assumption turns out to give the exactly known
critical frontier. 

\section{Evaluation of  $A$, $B$, $C$ for subnet networks}
\label{abc}
In this section we describe the computer algorithm we use to evaluate expressions of $A, B, C$ for
the Potts model with 2- and/or 3-site interactions in subnet networks.
  
\begin{figure}[b]
\includegraphics[scale=0.4]{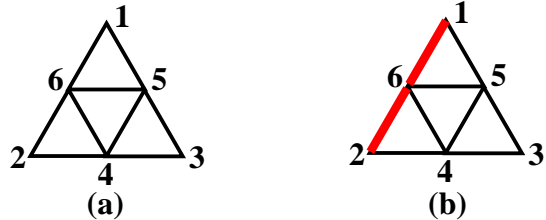}
\caption{(a) $2 \times 2$ subnet with pure 2-site interactions. (b) Red (bold) bonds are occupied, other bonds are vacant.}
\label{calabc}
\end{figure}

For the Potts model with 2-site interactions, consider the $2 \times 2$ subnet network in Fig. \ref{calabc}(a) 
as an example. The Boltzmann weight is
\begin{eqnarray}
W_{\triangle}=\sum\limits_{\{4,5,6\}}(1+v\delta_{1,6})(1+v\delta_{1,5})(1+v\delta_{2,4})\cdots \, ,
\label{exp22}
\end{eqnarray}
where $v=e^{K}-1$, $K$ is the 2-site coupling of the Potts model. 

Terms in the expansion of the products can be represented by graphs.
As shown in Fig. \ref{calabc}(a), there are $9$ bonds in the subnet. Define two states for each bond, occupied and vacant, 
then there are total $2^9=512$ graphs corresponding to the $512$ terms in the expansion of (\ref{exp22}). 
For example, Fig. \ref{calabc}(b) is a graph that corresponds to the 
term $\sum\limits_{\{4,5,6\}} v\delta_{1,6}v\delta_{2,6}=q^2v^2\delta_{1,2}$ 
contributing to $B$ with  a term $q^2v^2$. 

The 512 graphs are divided into five types according to following rules:

\begin{enumerate} 
\item Type-1, graphs with isolated spins 1, 2 and 3. The sum of these graphs
 generates the expression of $A$.
\item Type-2, graphs with spins 1 and 2 connected and spin 3 isolated. The sum of these 
graphs contributes to the expression of $B\delta_{12}$. For clarity we denote it as $B_{12}$.
\item Type-3, graphs with spins 2 and 3 connected and  spin 1 isolated. The sum of these 
graphs  contributes to the expression of $B\delta_{23}$ and denoted as $B_{23}$.
\item Type-4, graphs with spins 3 and 1 connected and spin 2 isolated. The sum of these 
graphs contributes to the expression of $B\delta_{31}$ and denoted as $B_{31}$. 
\item Type-5,  graphs with all  three spins 1, 2 and 3 connected. The sum of these graphs 
gives rise to the expression of $C$.
\end{enumerate}

For the Potts model with uniform and symmetric interactions, 
 we have $B_{12}=B_{23}=B_{31}=B$.

The algorithm of our program is to generate the $512$ graphs one by one, compute the weight of each graph,
and classify them into the five types. 
The graph weight assumes the form $q^{n_c}v^{n_v}$, where $n_v$ is the number of 
occupied bonds in the graph, and $n_c$ is the number of independent clusters 
isolated from, i.e., not connected to, sites 1, 2 or 3.
For example, the graph in Fig. \ref{calabc}(b) has $n_v=n_c=2$ and the weight $q^2v^2$.

The algorithm of our program is therefore as follows:

\begin{enumerate}
\item Generate one term, i.e., a graph, by choosing a set of occupied bonds.  
\item Count the number of independent clusters isolated from site 1, 2 or 3 as  $n_c$. 
\item Count the number of occupied bonds $n_v$.
\item Assign a term $q^{n_c}v^{n_v}$ to $A, B, $ or $C$ according to the aforementioned rules. 
\item Go to 1 for another graph until all 512 graphs are exhausted.
\end{enumerate}

\begin{figure}[b]
\includegraphics[scale=0.4]{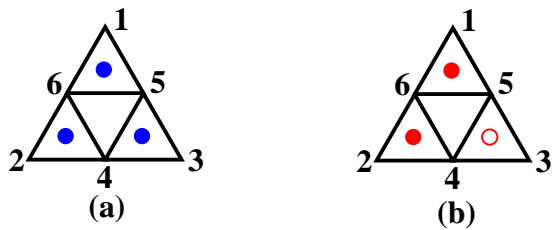}
\caption{(a) $2\times 2$ subnet with pure 3-site interactions indicated by dots. 
(b) Solid dot denotes the dot is occupied; open dot denotes it is vacant.}
\label{calabc1}
\end{figure}

 The procedure for the Potts model with pure 3-site interactions $M$
is similar. Take the case shown in Fig. \ref{calabc1}(a) as an example.
The doted up-pointing triangles have pure 3-site interactions
 and the Boltzmann weight of the $2 \times 2$ subnet can be written as 
\begin{equation}
W_{\triangle}=\sum\limits_{\{4,5,6\}}(1+m\delta_{1,5,6})(1+m\delta_{2,4,6})(1+m\delta_{3,4,5})\, , \label{exp22tri}
\end{equation}
where $m=e^M-1$.

To obtain an expansion in the form of (\ref{exp22tri}), we define two states of the dots as
either occupied or vacant. 
Thus there are   $2^3=8$ graphs corresponding to the 8 terms in  (\ref{exp22tri}). 
However, up to this point, clusters are defined by the connectivity of Potts sites, {\it not} by the dotted faces. But the 
connectivity can be readily translated to that of the dotted faces.
 A moment's reflections shows that 
the weight  contributing to $A$, $B$ or  $C$ is simply 
 $q^{n_c}m^{n_m}$, where $n_c$ is the number of independent clusters not containing
 sites 1, 2, or 3, and $n_m$ is the number of occupied dots. 

The rules to divide the graphs into five types corresponding to $A,B,C$ are the same
as the ones  for pure 2-site interactions.
For example, the graph in Fig. \ref{calabc1}(b) has 
$n_m=2$, $n_c=0$ and  corresponds to the term
$\sum\limits_{4, 5, 6} (m\delta_{1,5,6})(m\delta_{2,4,6})=m^2 \delta_{12}$, thus 
contributing to $B_{12}$ with a term $m^2$. 

The algorithm to obtain expressions of $A,B,C$  is therefore very 
similar to the one described in the above for  2-site interactions:

\begin{enumerate}
\item Generate one  term, i.e.,  a graph, by choosing a set of occupied dots.
\item Count the number of  clusters isolated from sites 1, 2 or  3 as  $n_c$.
\item Count the number of occupied dots $n_m$.
\item Assign $q^{n_c}m^{n_m}$ to $A, B$ or $C$ respectively according to the aforementioned rules.
\item Go to 1 for another graph until all possible graphs are exhausted. 
\end{enumerate}

In the Appendix we present expressions of $A,B,C$ for the Potts model on  $n\times n$ subnets  with 2-site 
interactions for $n \leq 4$,  and for subnets with 3-site interactions for $n \leq 7$.
 
\section{The transfer matrix and finite-size scaling}
\label{TM}
\begin{figure}
\includegraphics[scale=0.4]{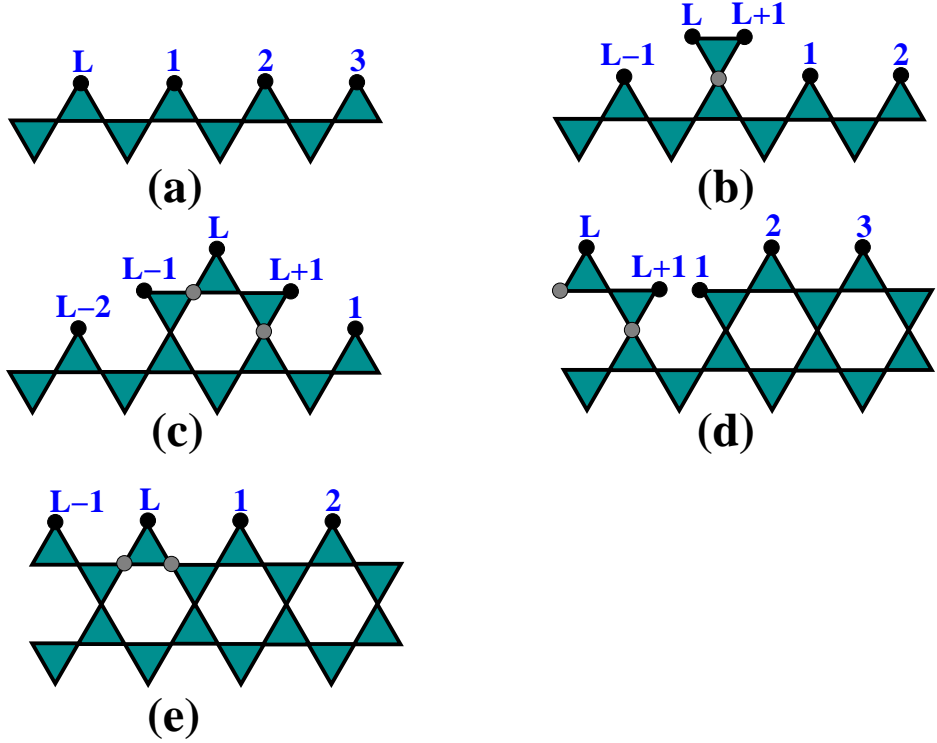}
\caption{Construction of the transfer matrix for kagome-type subnet lattices.
The procedure consists of several steps each corresponding  to a sparse transfer matrix.}
\label{subnettm}
\end{figure}

We use the method of transfer matrix  to calculate statistical variables for lattice models 
wrapped on a cylinder with circumference $L$ and length $N$. For lattices shown in Fig. \ref{tri&kag}
with hatched triangles, 
$L$ and $N$ count 
up- and down-pointing hatched triangles (rather than individual Potts spins within each triangle).
Thus, for an $(m\times m):(n\times n)$ lattice
 shown in Fig.  \ref{subnettm}(a), there are actually
$(m+n)L$ Potts spins in a length $L$. 
 
For the Potts model, we build the transfer matrix by using the random-cluster representation 
of the Potts partition function
\cite{FK1,FK2} 
\begin{equation}
Z= \sum_{g}v^{n_b(g)}q^{n_c(g)}  \label{wtri},
\end{equation}
where the summation is over all subgraphs $g$ of the lattice (or graph)  on which
the Potts model is defined, 
 $v=e^K-1$, and $n_b(g) $ and $n_c(g)$ are, respectively, the  number of bonds and clusters in $g$.
For the ($q=1$) bond percolation we have simply $Z=(1+v)^{E}$, where $E$ is the total number of 
edges of the lattice.
 
The concept of connectivity plays an essential role in the building of the transfer matrix. 
Sites that belong to the same cluster are said to be connected. 
In the $N \times L$ cylinder,
  each of the $L$ end sites of the cylinder is either isolated from or connected to 
other end sites. 
The connectivity of the $L$ end sites of the cylinder is described by non-crossing partitions
of the  sites. There are a total of
\be
d_L= {\frac 1 {2L+1}} {{2L+1} \choose  {L}}
\ee
 such non-crossing partitions \cite{temperleylieb}  indexed by
$\beta$. In the transfer matrix consideration
the non-crossing partitions are mapped onto and coded by a set of integers
$1, 2, \cdots, d_L$.
 A detailed explanation of the coding procedure 
can be found in  \cite{pottsTM}.

The partition function of the Potts model 
can therefore be written as 
\begin{equation}
Z^{(N)}=\sum\limits_{\beta} Z^{(N)}_\beta,
\end{equation}
where $Z^{(N)}_\beta$ is the partition sum restricted to the partition $\beta$.
 The restricted sums  $Z^{(N)}_\alpha$ and $Z^{(N-1)}_\beta$ are connected by a transfer matrix $T$ in the form of
 \begin{equation}
Z^{(N)}_\alpha=\sum\limits_{\beta} T_{\alpha, \beta}Z^{(N-1)}_\beta\, ,
\end{equation}
where
\begin{equation}
T_{\alpha, \beta}=\sum\limits_{g} v^{\Delta n_b(g)}q^{\Delta n_c(g)} 
\label{tmweight}
\end{equation}
are elements of $T$. Clearly, $T$ has the dimension $d_L \times d_L$.
It is also clear that the summation in (\ref{tmweight}) 
is over subgraphs $g$ connecting partitions $\beta$ and $\alpha$ of the $(N-1)$-th and $N$-th rows, with
 $\Delta n_b(g)$ and $\Delta n_c(g)$ denoting, respectively, the net (positive or negative) change of the
number of bonds  and clusters due to the introduction of $g$. 

To conserve computer memory and running time, the transfer matrix is 
converted  into a product of sparse matrices as described below 
(see \cite{pottsTM} for further details). 
This technique has  proved to be very efficient in the transfer
matrix study of the Potts model, 
the O($n$) loop  and other lattice models \cite{OnTM, qianxiaofeng, wenan1, wenan2,wenan3, wenan4, wenan5}.

The transfer matrix can  be regarded as adding a new layer to the system. This process 
converts  the transfer matrix $T$  into a product of $L+1$ sparse matrices for kagome-type lattices 
of Fig. \ref{tri&kag}(b). 
The first sparse matrix $T_1$ adds one down-triangle with two `new' corner sites on  top of 
an `old' site followed by a shift of  labeling of  sites. This gives rise to a new  layer with $L+1$ sites as 
shown in Figs. \ref{subnettm}(a) and \ref{subnettm}(b).  The matrix $T_1$ is a $d_{L+1} \times d_{L}$ 
rectangular matrix. 
The second sparse matrix $T_2$ adds one up- and one down-triangle simultaneously. 
By shifting the labels of the sites of the top (new) layer cyclically, 
$T_2$ brings in two new  corner sites $L$  and $L+1$  
on  top of the current layer, and covers two old  corner sites $L+1$  and $1$ , as shown in 
Figs. \ref{subnettm}(b) and \ref{subnettm}(c).  
$T_2$ is a $d_{L+1} \times d_{L+1}$ square matrix. 
After $L-1$ such steps,  the graph is transformed to the one shown in Fig. \ref{subnettm}(d).
By adding an up-triangle on the two old  sites $L+1$  and $1$  under  cylindrical boundary condition,  
the last sparse matrix $T_{3}$ adds the last new  corner site $L$  to the system  
as shown from  Figs. \ref{subnettm}(d) and  \ref{subnettm}(e). Labels of the top sites
are shifted and $T_3$ is a $d_L \times d_{L+1}$ rectangular matrix.
 
The transfer matrix $T$  now assumes the form of a product of sparse matrices,
\begin{equation}
T=T_3 T_2^{L-1}T_1\, .
\end{equation}
In the actual calculation, we need to store only the positions and  values of the nonzero elements of each sparse matrix in a
few one-dimensional arrays. 

 In constructing these sparse matrices, one needs to enumerate all possible graphs inside the added 
(one or two) hatched triangles for a given connectedness of the partitions of the new and  old layers. 
For example, in the construction of $T_1$,  we need to add  a subnet shown in Fig. \ref{subnets} (flipped vertically). 
In the case of $1\times 1$ subnet, it is straightforward to enumerate all  graphs manually.
However, in the case of $2\times 2$ and  higher order subnets, 
it is tedious and sometimes impossible to count all possible graphs by hand. Therefore, we  make use of a
computer algorithm 
similar to the one used in obtaining expressions of $A, B, C$ to count $\Delta n_b$ and $\Delta n_c$.  
 
For site percolation, the partition function is
\bea
Z&=& \sum_{g}s^{n_s(g)} (1-s)^{N-n_s(g)} \nonumber \\
 &=& 1\, , 
\label{Zsite}
\eea
where the summation is over all site percolation configurations $g$, 
$s$ is the probability that a site is occupied, $n_s(g) $ is the number of occupied sites in $g$, 
and $N$ is the
total number of sites. 
The corresponding transfer matrix is defined in a way similar to that of the 
Potts model, but with a twist due to the
presence of vacant sites and henceforth not all $L$ end sites are occupied.  
Denote the number of end sites  that are occupied by $n=0,1,\cdots,L$ which can be distributed
in ${L \choose n}$ different ways. 
 Then there are a total of
\be
d^S_L= \sum_{n=0}^L {L \choose n} d_n
\label{dsL}
\ee
non-crossing partitions and the transfer matrix has the dimension $d^S_L\times d^S_L$. 
It is clear that we have $d^S_L > d_L$.
 
The partitions  can again be coded by means of a sequence of integers $1,2,\cdots, d^S_L$.
The coding algorithm is the same as that used in the consideration of the Potts model with vacancies 
\cite{qianxiaofeng} and in the  study of site percolation  \cite{kagomefeng}.

To determine the critical threshold of the Potts model and/or site percolation, we 
calculate the magnetic scaled gap
\begin{equation}
 X_h(v,L)=\frac{L}{2\pi\xi_h(v,L)}\, ,
\label{Xh}
\end{equation}
where $\xi_h(v,L)$ is the magnetic correlation length (with $v$  replaced by $s$
for site percolation).
In the language of the random cluster model and site percolation, the magnetic correlation function is 
defined to be the probability that two sites at a distance  $r$ belong to the same cluster, or
\begin{equation}
g_{r}=\frac{Z^{\prime}}{Z} \, ,
\end{equation}
where $Z^{\prime}=\sum_{g} ' v^{n_b(g)}q^{n_c(g)}$ for the random cluster model 
and $Z^{\prime}=\sum_{g} ' s^{n_s(g)} (1-s)^{N-n_s(g)}$ for site percolation. The summations
in $Z'$ are  the same  as in (\ref{wtri}) and (\ref{Zsite}) but restricted to
subgraphs $g$  
with at least one cluster spanning from row 1 to row $r$.

We  define a transfer matrix, hereafter referred to as the magnetic sector of the transfer matrix, based on $Z^{\prime}$, 
in a way  similar to that of the transfer matrix based on $Z$ in (\ref{tmweight}) or (\ref{Zsite})
in the 'non-magnetic' sector. 
 In constructing  the magnetic sector of the transfer matrix,
we use the `magnetic' type connectivity of the $L$ end sites of the cylinder, which, in addition
to describing how sites are connected,
specifies which sites are still connected to a site in row 1.
These sites are called `magnetic sites'.
To count the total number of  non-crossing partitions describing the `magnetic' type
connectivity,
we first code the positions of the magnetic sites by means of a binary
number $m=0,1,...,2^L-1$, where the binary digit 1 denotes a magnetic site. 
The magnetic sites divide the remaining sites in $g(m)$ groups such that
two sites in different groups cannot be connected. Let $n(j)$ be the number of sites in the 
$j$-th group. 
Then there are 
\be
h_m=d_{n(1)}d_{n(2)}\cdots d_{n(g(m))}
\ee
non-crossing partitions for the Potts model and 
\be
h^S_m=d^S_{n(1)}d^S_{n(2)}\cdots d^S_{n(g(m))}\,
\ee
non-crossing partitions for site percolation.
The total number of non-crossing partitions is therefore
 \be
d^{(m)}_L=\sum_{m=0}^{2^L-1} h_m
\ee
for the Potts model, and
\be
d^{S(m)}_L=\sum_{m=0}^{2^L-1} h^S_m
\ee
for the site percolation. 

The partitions  can again be coded by means of a sequence of integers.
A detailed description of the coding algorithm can be found in \cite{qianxiaofeng}. 
The magnetic sectors of the transfer matrix now have the dimensions 
$d^{(m)}_L\times d^{(m)}_L$ and $d^{S(m)}_L\times d^{S(m)}_L$ for the Potts model
and site percolation, respectively,
and are much larger than those of the non-magnetic sectors. 
The magnetic sector of the transfer matrix 
can also be converted into a product of sparse matrices in the same way as 
in the case of the non-magnetic sector.

The inverse magnetic correlation length is given by
\begin{equation}
\frac{1}{\xi_h(v,L)}= \zeta \ln \bigg( \frac{\lambda_0}{\lambda^{\prime}_0}\bigg) \, ,
\end{equation}
where $\lambda_0$ and $\lambda_0^{\prime}$ are the leading eigenvalues of the transfer matrix in the 
non-magnetic and magnetic sector respectively, $\zeta$ is a geometrical factor which is the ratio
between the unit of $L$ and the thickness of a layer added by the transfer matrix.
The magnetic scaled gap then follows.
 
According to  finite-size scaling theory \cite{fss}  and Cardy's conformal mapping \cite{conformalmapping}, 
$X_h(v,L)$ can be expanded as
\begin{equation}
X_h(v,L)=X_h+atL^{y_t}+buL^{y_u}+...\label{Xhfs} \, ,
\end{equation}
where $X_h$ is the magnetic scaling dimension,
$t$ is the deviation from the critical point, 
and $u$ the irrelevant field. Here,
$y_t$ is the thermal renormalization exponent, $y_u$ the leading irrelevant renormalization exponent, and  
$a$ and $b$ are unknown constants. 

We substitute (\ref{Xhfs}) into the finite-size scaling equation 
connecting  lattices  of sizes  $L$ and $L-1$,
\begin{equation}
 X_h(v,L)=X_h(v,L-1)\, ,
\label{XhL}
\end{equation}
and denote the solution of (\ref{XhL}) by $v_c(L)$, which has the expansion
\begin{eqnarray}
v_c(L) =v_c + a' u L^{y_u-y_t}+\cdots\,  ,
\label{vc}
\end{eqnarray}
where $a'$ is an unknown constant. Because $y_u<0$ and $y_t>0$,
$v_c(L)$ for a sequence of increasing system sizes converge to the critical point $v_c$.

At $v_c(L)$, the expression $X_h(v_c(L),L)$ for a sequence of sizes $L$ converge to the magnetic
scaling dimension $X_h$ as
\be
X_h(v_c(L),L)=X_h+b' u L^{y_u}+... \, ,
\label{Xhc}
\ee
with $b'$ an unknown constant. This  determines the magnetic scaling dimension $X_h$.

The free energy per unit distance is given by
\begin{equation}
f(L)=\frac{\zeta \ln \lambda_0}{L} \, , \label{freeenergydensity}
\end{equation}
where $\lambda_0$ is the largest eigenvalue of the transfer matrix in the non-magnetic sector.
According to conformal invariance theory, the large-$L$ asymptotic finite-size
behavior of the free energy density at the critical point is \cite{BCN, Affleck}
\be
f(L) \simeq f(\infty) +\frac{\pi c}{6 L^2}\, ,
\label{cc}
\ee
where $c$ is the conformal anomaly.

The conformal anomaly $c$ and the magnetic scaling dimension $X_h$ are two important universal quantities defining the
universality class. 
 For the two-dimensional $q$-state Potts model,
they are given by the conformal invariance theory and Coulomb gas method 
\cite{BCN,Affleck,dNijs, BE,cg} as
\be
c=1-\frac{6(1-g)^2}{g},~~~~ X_h=1-\frac{g}{2}-\frac{3}{8g}\, ,
\ee
where
\be
\sqrt {q}=-2 \cos(\pi g), ~~~ \frac{1}{2}\leq g \leq 1 \, .
\label{qg}
\ee

\section{Numerical and some exact results}
\label{list_result}
In this section we present numerical results of our transfer matrix calculations 
and finite-size scaling analysis for the 3-12 and kagome-type lattices. 
We also present some exact results for site percolation on the $(1\times 1):(n\times n)$ lattices.

\subsection{The $q$-state Potts model on the 3-12  and 
kagome-type subnet lattices}
\label{pottsbond}
 Critical points are estimated by extrapolating the solutions of  (\ref{XhL}) 
for a sequence of increasing system sizes in accordance with the finite-size scaling equation (\ref{vc}). 
The numerical accuracy one reaches depends highly on the system size reached in the calculation.

For the 3-12 lattice and the $(1\times 1):(1\times 1)$ (the  kagome) and $(2\times 2):(2\times 2)$ kagome-type 
subnet lattices,   the largest 
dimension of the arrays used to store the values and the positions of nonzero elements of the sparse matrices 
is $d^{(m)}_{L+1}$.  The largest system size we reached is $L=15$ with
$d^{(m)}_{16}=335897865$.  
For the $(3\times 3):(3\times 3)$ and $(4\times 4):(4\times 4) $ kagome-type subnet lattices, the sparse 
matrix $T_2$ in the magnetic sector is further decomposed in two rectangular matrices of dimensions 
$d^{(m)}_{L+1} \times d^{(m)}_{L+2}$  and $d^{(m)}_{L+2} \times d^{(m)}_{L+1}$, 
and the largest system size we reached is  $L=14$. 
The computer memory requirement for the calculations of the largest system is about 65 gigabytes, which is quite large, but
the CPU time consumed is rather modest. It is just a few hours for a typical calculation of the magnetic scaled gap.

The magnetic scaling dimension $X_h$
is estimated by extrapolating the scaled gaps $X_h(v_c(L),L)$ at the solution of  (\ref{XhL}) for
a sequence of increasing system sizes in accordance with (\ref{Xhc}). 
The free energy density at the estimated critical point is calculated using (\ref{freeenergydensity})
and the conformal anomaly $c$ is computed by making use of the finite-size scaling relation (\ref{cc}).
Details of the data fitting procedure are described in \cite{pottsTM}. 
We also checked corrections to scaling due to the leading irrelevant field. 
Take the simple kagome lattice as an example. 
According to the Coulomb gas  theory \cite{cg},  $y_{t_2}=4-4/g$ with $g$ given in (\ref{qg}) is the second
leading thermal exponent, which we expect to be a candidate for the leading correction exponent $y_u$. 
For $q > 2$, we indeed found $y_u$ close to $y_{t_2}$. 
For $q=0.5$, $y_u$ is about $-2.00(1)$, which is the analytic one. For $q=1.0$ and 1.5, we found
$y_u=-1.79(3)$ and $-1.51(2)$ respectively, which dominate and overcome the corresponding $y_{t_2}$. 
For $q=2$,  the Ising model, we obtain $y_u= -4.00(1)$. The amplitudes of $y_{t_2}=-4/3$ and the analytic
$y_u=-2$ corrections vanish.  This  is understandable for lattices with sixfold rotational symmetry. 
This picture is generally true for all $(n\times n):(n\times n)$ kagome subnet lattices.

\begin{table}[htbp]
\caption{Critical properties of the Potts model on the 3-12 lattice. (H $=$ Homogeneity Assumption,
N $=$ Numerical, T = Theoretical universality prediction.)}
 \begin{tabular}{c|l|l|l|l|l|l}
\hline
      $q$ &$v_c$\quad (H) &  $v_c$\quad (N)& c\quad (T)& c\quad (N)& $X_h$\quad (T)&$X_h$\quad (N) \\
    \hline
    $0.5$     &2.007916417382387  &2.00788(1)  &-0.445833945 &-0.4458340(1)&0.082757037 &0.08276(1)\\
    $1.0$     &2.852426157798754  &2.8523883(2) &0           &0            &0.104166667 &0.104167(1) \\
    $1.5$     &3.510849695265078  &3.510825(2) &0.288024142  &0.288024(1)  &0.116778423 &0.116778(1) \\
    $2.0$     &4.073446135573680  &4.0734460(1)&0.5          &0.500000(1)  &1/8         &0.12500000(1) \\
    $2.5$     &4.574927577671523  &4.574952(3) &0.66584083   &0.66585(1)   &0.130338138 &0.13033(1) \\
    $3.0$     &5.033022514872745  &5.033077(3) &4/5          &0.800(1)     &2/15        &0.13333(1) \\
    $3.5$     &5.458234413883058  &5.458313(2) &0.910294591  &0.91(1)      &0.133771753 &0.1339(3) \\
    $4.0$     &5.857394827983647  &5.857497(3) &1            &0.999(1)     &1/8         &0.13(1) \\

    \hline
\end{tabular}
\label{pc312}
\end{table}

\begin{table}[htbp]
\caption{Critical properties of the Potts model on the kagome ($1\times 1$) lattice. 
(H $=$ Homogeneity Assumption,
N $=$ Numerical, T $=$ Theoretical universality prediction.)}
 \begin{tabular}{c|l|l|l|l|l|l}
    \hline
      $q$   &$v_c$\quad(H) &  $v_c$\quad(N) &$c$\quad(T) &$c$\quad(N)& $X_h$\quad(T) &$X_h$\quad(N)\\
    \hline
    $0.5$     &0.787417375457453 &0.787320(1) &-0.445833945 &-0.445834(1)  &0.082757037      &0.082757(1)\\
    $1.0$     &1.102738621067509 &1.10262924(2) &0           &0             &0.104166667      &0.104167(1) \\
    $1.5$     &1.342082948593078 &1.3420126(2) &0.288024142 &0.2880243(3)  & 0.116778423     &0.116780(3)\\
    $2.0$     &1.542459756837412 &1.5424598(1) &0.5         &0.500000(1)  &1/8               &0.12500000(1)\\
    $2.5$     &1.718102046569530 &1.718191(3)  &0.66584083  &0.66584(1)    &0.130338138      &0.1304(1)\\
    $3.0$     &1.876269208345760 &1.876458(3) &4/5         &0.8000(1)     &2/15             &0.1333(1)\\
    $3.5$     &2.021253955272383 &2.02154(2)  &0.910294591 &0.910(1)      &0.133771753      &0.134(1)\\
    $4.0$     &2.155842236513638 &2.15620(5)   &1           &1.00(1)       &1/8              &0.13(1)\\
    \hline

\end{tabular}
\label{pckagome}
\end{table}

\begin{table}[htbp]
\caption{Critical properties of the Potts model on the $(2 \times 2):(2 \times 2)$ kagome lattice.
(H $=$ Homogeneity Assumption,
N $=$ Numerical, T $=$ Theoretical universality prediction.)}
\begin{tabular}{c|l|l|l|l|l|l}
    \hline
      $q$     &$v_c$\quad (H)  &  $v_c$\quad(N)  &$c$\quad(T) &$c$\quad (N) &$X_h$\quad(T) &$X_h$\quad(N)\\
    \hline
    $0.5$     &1.115482279992555 &1.1154309(3) &-0.445833945 &-0.4458340(2)&0.082757037 &0.082757(1) \\
    $1.0$     &1.505450910604828 &1.5053987(1)&0            &0            &0.104166667 &0.104167(1) \\
    $1.5$     &1.790803965420646 &1.7907720(2) &0.288024142  &0.288024(1)  &0.116778423 &0.11678(1)\\
    $2.0$     &2.024382957091806 &2.02438295(3)&0.5          &0.500000(1)  &1/8         &0.1250000(1)\\
    $2.5$     &2.225885325024986 &2.2259229(2) &0.66584083   &0.66584(1)   &0.130338138 &0.13034(1)\\
    $3.0$     &2.405138877193783 &2.4052181(3)&4/5          &0.8001(1)    &2/15        &0.1333(1) \\
    $3.5$     &2.567855953492942 &2.567981(2) &0.910294591  &0.910(1)     &0.133771753&0.1339(3)\\
    $4.0$     &2.717691692682905 &2.717856(2) &1            &0.99(1)      &1/8         &0.13(1) \\
    \hline
\end{tabular}
\label{pckagome2}
\end{table}

\begin{table}[htbp]
\caption{Critical properties of the Potts model on the $(3 \times 3):(3 \times 3)$ kagome lattice. 
(H $=$ Homogeneity Assumption,
N $=$ Numerical, T $=$ Theoretical universality prediction.)}
  
\begin{tabular}{c|l|l|l|l|l|l}
    \hline
      $q$ &$v_c$\quad(H) &  $v_c$\quad(N) &$c$\quad(T) &$c$\quad(N) &$X_h$\quad(T) &$X_h$\quad(N)\\
    \hline
    $0.5$     &1.236699591471530 & 1.2366855(3) &-0.445833945&-0.4458340(2)  &0.082757037  &0.0827569(2)\\
    $1.0$     &1.626971272019731 & 1.6269594(2) &0           &0             &0.104166667  &0.104167(1)\\
    $1.5$     &1.906766682469675 & 1.906760(1) &0.288024142 &0.288024(1)   &0.116778423  &0.116779(2)\\
    $2.0$     &2.133002727374153 & 2.13300273(1)&0.5         &0.500000(1)   &1/8          &0.1250000(1)\\
    $2.5$     &2.326449318777172 & 2.32645568(5) &0.66584083  &0.66585(1)    &0.130338138  &0.130338(1)\\
    $3.0$     &2.497336478778200 & 2.4973486(2) &4/5         &0.800(1)      &2/15         &0.1333(1)\\
    $3.5$     &2.651556985414795 & 2.651575(3) &0.910294591 &0.91(1)       &0.133771753  &0.1338(1)\\
    $4.0$     &2.79285603450327  & 2.79288(2)  &1           &0.999(1)      &1/8          &0.13(1)\\
    \hline
\end{tabular}
\label{pckagome3}
\end{table}

\begin{table}[b]
\caption{Critical properties of the Potts model on the $(4 \times 4):(4 \times 4)$ kagome lattice. 
(H $=$ Homogeneity Assumption,
N $=$ Numerical, T $=$ Theoretical universality prediction.)}
 
\begin{tabular}{c|l|l|l|l|l|l}
    \hline
      $q$     &$v_c$\quad(H) &  $v_c$\quad(N) &$c$\quad(T) &  $c$\quad(N) &$X_h$\quad(T)&$X_h$\quad(N)\\
    \hline
    $0.5$     &1.287715536704650   &     1.2877116(2) &-0.445833945 &-0.4458340(1) &0.082757037&0.0827569(1)\\
    $1.0$     &1.669262339202358   &     1.6692593(3) &0            &0             &0.104166667&0.10417(1)\\
    $1.5$     &1.941284616762751   &     1.9412832(5) &0.288024142  &0.288024(1)   &0.116778423&0.11678(1) \\
    $2.0$     &2.160721132019555    &    2.160721132(1)&0.5          &0.500000(1)   &1/8        &0.1250000(1)  \\
    $2.5$     &2.348099505779181   &     2.3481001(2) &0.66584083   &0.66585(1)    &0.130338138&0.13034(1)  \\
    $3.0$     &2.513467694176093   &     2.5134684(2) &4/5          &0.800(1)      &2/15       &0.1333(1) \\
    $3.5$     &2.662592230189568   &     2.662594(3)  &0.910294591  &0.911(1)      &0.133771753&0.134(1) \\
    $4.0$     &2.799129506399588   &     2.799132(5)  &1            &0.999(1)      &1/8        &0.13(1) \\
    \hline
\end{tabular}

\label{pckagome4}
\end{table}

\begin{table}[b]
\caption{The thresholds $p_c$ of bond percolation on the 3-12 and various kagome-type subnet lattices. 
(H $=$ Homogeneity Assumption, N $=$ Numerical.)}
\begin{tabular}{c|l|l|l}
    \hline
      Subnet           &$p_c$\quad(H)          &  $p_c$\quad(N)    & Other sources\\
    \hline
$(1\times 1):(1\times 1)$     &0.524429717521274 &0.524404978(5) &0.52440499(2) \cite{kagomefeng}\\
& & &0.52440503(5) \cite{kagomefeng}\\
& & &0.5244053(3) \cite{ZiffSudingKagome}\\
$(2\times 2):(2\times 2)$     &0.600870248238631 &0.60086193(3) &0.6008624(10) \cite{kagomesubnets}\\
$(3\times 3):(3\times 3)$     &0.619333484666866 &0.61933176(5) &0.6193296(10) \cite{kagomesubnets}\\
$(4\times 4):(4\times 4)$     &0.625364661497144 &0.62536424(7) &0.625365(3) \cite{kagomesubnets}\\
    \hline
3-12 lattice                  &0.740423317919897 &0.74042077(2)  &0.74042118 \cite{ScullardZiffbondkagome312}\\
    &&&0.74042081 \cite{kagomesubnets}\\
    &&&0.74042195(80) \cite{bond312}\\
    \hline
\end{tabular}
\label{bondpc}
\end{table}

We summarize in   Tables \ref{pc312}-\ref{pckagome4} 
numerical results of our calculations
on the critical point $v_c$, conformal anomaly $c$, and magnetic scaling dimension $X_h$
 together with  the universality predictions of $c$ and $X_h$.
We have also computed $v_c$  using the homogeneity assumption and list the results.
The $v_c$ calculation for the kagome lattice extends those of \cite{kagomepotts} using Monte Carlo 
renormalization group method and finite-site scaling analysis  for $q=1, 2, 3, 4.$ 
Our study extends to non-integer $q$  and offers results with higher accuracy.
 
For  $q=2$, the Ising model, our numerical estimates of the critical threshold 
 agree with the exact critical results up to $7$ or $8$ decimal numbers.
 This probably indicates
the limit of the numerical accuracy of the finite-size analysis 
we can reach at present.
For  $q \neq 2$, the critical points obtained from (\ref{conj}) under the homogeneity assumption  coincide
with our numerical estimations to 5 or so decimal places but lie outside error bars.  
This indicates that the homogeneity assumption, while highly accurate,  
is an excellent approximation yielding numerical values with an error within one part in $10^5$.
Our computed values of the conformal anomaly $c$ and the magnetic 
scaling dimension $X_h$  coincide with the theoretical universality predictions within  error bars. 

Our numerical results for bond percolation are summarized in Table \ref{bondpc}
for the 3-12 lattice and the $(n\times n):(n\times n)$ subnet kagome-type lattices. 
For the kagome lattice, we found $p_c=0.524404978(5)$, which coincides with the best estimation \cite{kagomefeng}.
For  the 3-12 lattice 
our numerical result of $p_c=0.74042077(2)$  is in agreement with other  
findings \cite{ScullardZiffbondkagome312,kagomesubnets,bond312} to 6 decimal places.
For kagome-type subnet lattices, our numerical analysis determines $p_c$ with an accuracy up to
7 or 8 decimal places. 

In Table \ref{bondpc}
we also give thresholds computed using the homogeneity assumption (\ref{conj}).
 The polynomial equations determining the bond percolation thresholds $p_c$ under the
homogeneity assumption (\ref{conj}) for  $(n\times n):(n\times n)$ subnet lattices
in Table  \ref{bondpc} are as follows: 
\bea
&&1-3p^2-6p^3+12p^4-6p^5+p^6= 0, \quad (n=1) , \label{11HA} \\
&& 1-3p^4-18p^5-39p^6+30p^7+273p^8+264p^9-1785p^{10}- 126p^{11}+8232p^{12}\nonumber \\ 
&& -162326p^{13} + 16359 p^{14} -9948p^{15}+3708p^{16} - 786p^{17} +73p^{18} = 0,\quad (n=2), \label{22HA}
\eea
\begin{eqnarray}
1&-&3p^6-36p^7-186p^8-372p^9+447p^{10}+3558p^{11}+4711p^{12}-5274p^{13}-30771p^{14}\nonumber\\
&-&110816p^{15}+69828p^{16}+1309302p^{17}-242760p^{18}-10117626p^{19}+9190737p^{20}\nonumber\\
&+&53446600p^{21}-137597577p^{22}-15101358p^{23}+714425889p^{24}-1897059306p^{25}\nonumber\\
&+&2985201585p^{26}-3337272356p^{27}+2817156177p^{28}-1840940730p^{29}\nonumber\\
&+&938230487p^{30}-371179194p^{31}+112125462p^{32}-25052124p^{33}+3909120p^{34}\nonumber\\
&-&380880p^{35}+17464p^{36}=0,\quad (n=3)
\end{eqnarray}
\begin{eqnarray}
1&-&3p^8-60p^9-528p^{10}-2406p^{11}-4518p^{12}+8388p^{13}+64323p^{14}+108744p^{15}-149520p^{16}\nonumber \\
&-&892404p^{17}-664532p^{18}+2272086p^{19}-2348817p^{20}-12425874p^{21}+123063933p^{22}\nonumber\\
&+&344663478p^{23}-1382031989p^{24}-5244471786p^{25}+12598666671p^{26}+50539880448p^{27}\nonumber\\
&-&112896871341p^{28}-350902330710p^{29}+955575283123p^{30}+1782743557128p^{31}\nonumber\\
&-&7239409905561p^{32}-5767231526534p^{33}+52365034246041p^{34}-23401813013430p^{35}\nonumber\\
&-&353073527306441p^{36}+1041090144149322p^{37}-623756767891383p^{38}-4367477247915326p^{39}\nonumber\\
&+&18117607172859264p^{40}-42034422047996604p^{41}+71675099615055545p^{42}\nonumber\\
&-&97479081216503664p^{43}+109775262989475858p^{44}-104474772230850020p^{45}\nonumber\\
&+&85036825023972936p^{46}-59604466077733650p^{47}+36101308809040333p^{48}\nonumber\\
&-&18909591474961260p^{49}+8552494666923729p^{50}-3327421649714158p^{51}+1106659102637175p^{52}\nonumber\\
&-&311767535257674p^{53}+73442507365712p^{54}-14206464131418p^{55}+2198697552561p^{56}\nonumber\\
&-&261883431344p^{57}+22544948382p^{58}-1248899580p^{59}+33437353p^{60}=0,  \quad (n=4). \label{44HA}
\end{eqnarray}
The threshold (\ref{11HA}) for $n=1$ has previously been given in \cite{wu79,kagomepotts} 
 and in  \cite{ScullardZiffbondkagome312,SZ}.
Thresholds (\ref{22HA}) - (\ref{44HA}) for $n=2,3,4$ are new. 
The polynomial equation determining $p_c$ under the homogeneity assumption in Table \ref{bondpc} 
for the 3-12 lattice  has been given in I and \cite{SZ}, and is \ 
$1-p+p^2+p^3-7p^4+4p^5=0$.

\subsection{Site percolation on the 3-12 lattice}
The exact critical threshold for site percolation on the 3-12 lattice
is  known to be $s_c=\sqrt{1-2\sin(\pi/18)}$.
It  was first given in \cite{suding1999} and is shown in  I to be
the same as that of
 the $(2 \times 2):(2 \times 2)$ Potts subnet lattice with pure 3-site interactions.
To calibrate our numerical approach, we have also computed $s_c$ using the transfer matrix approach.
 Our numerical determination of critical properties of site percolation is
 summarized in the last row in Table \ref{pcsite}. The comparison of numerical estimates  of thresholds
with exact results shows  agreements  up to 7 decimal places, indicating
 our numerical estimates to be accurate to the same degree of accuracy.
Our numerical determination of the conformal anomaly and magnetic scaling dimension of site percolation
indicates that these models all belong to the two-dimensional $q=1$ Potts model universality class. Again, the 
hypothesis of universality is verified.

\begin{table}[h]
\caption{Critical properties of site percolation on $(1 \times 1):(n\times n)$ kagome-type subnet lattices and 
the 3-12 lattice.  
(E $=$ Exact result, N $=$ Numerical, T $=$ Theoretical universality prediction.)}
\begin{tabular}{c|l|l|l|l|l|l|l}
    \hline
    Subnet                        &  $s_c$\quad(E)    &$s_c$\quad(N)& Other sources &$c$(T) &$c$(N) &$X_h$\quad(T)&$X_h$\quad(N) \\
    \hline
    $(1\times 1):(1\times 1)$     & 0.652703644666139 & 0.6527035(2)& $1-2\sin (\pi/18)$ \cite{kagomesite} &0&0&0.1041667&0.1042(1)\\
    $(1\times 1):(2\times 2)$     & 0.707106781186548 & 0.7071068(2)& $1/\sqrt{2}$ \cite{akagome21}&0&0          &0.1041667&0.10416(1)\\
    $(1\times 1):(3\times 3)$     & 0.728355596425196 & 0.7283555(1)&               &0                           &0&0.1041667&0.10417(1)\\
    $(1\times 1):(4\times 4)$     & 0.738348473943256 & 0.7383483(5)&               &0                           &0&0.1041667&0.10417(2)\\
    $(1\times 1):(5\times 5)$     & 0.743548682503071 & 0.7435486(3)&               &0                           &0&0.1041667&0.1042(1)\\
    $(1\times 1):(6\times 6)$     & 0.746418147634282 & 0.7464180(3)&               &0                           &0&0.1041667&0.10417(1)\\
    \hline
   3-12  lattice                  & $\sqrt{1-2\sin (\pi/18)}$& 0.8079008(3)&$\sqrt{1-2\sin (\pi/18)}$ \cite{suding1999} &0&0&0.1041667&0.10416(1)\\
                    & =0.807900764120 &             &0.807904(4) \quad [N] \cite{suding1999}&    & &   &\\
    \hline

\end{tabular}
\label{pcsite}
\end{table}

\subsection{Exact thresholds for site percolation on the $(1\times1):(n\times n)$ kagome-type lattice}
It was shown in I, and in
Sec. \ref{frontiers}, that the rigorous critical frontier (\ref{CA}) yields the exact thresholds of site percolation on 
$(1\times1):(n\times n)$ kagome-type subnet lattices.
 The polynomial equations determining the threshold
$s_c$ are generated by substituting expressions of $A$ and $C$ in (A.5) - (A.11) into 
(\ref{CA}) and setting $q=1$. This yields the site percolation thresholds shown in Table \ref{pcsite}.
Explicitly, the thresholds for site
percolation on  $(1\times1):(n\times n)$ kagome-type subnet lattices, $1\leq n \leq 6$, 
in Table \ref{pcsite} are as follows:
\bea
&& 1-3s^2+s^3=0, \quad (n=1) \\
&& 1-3s^3-3s^4+6s^5-2s^6=0, \quad (n=2) \\
&& 1-3s^4-9s^5+9s^6+11s^7-12s^8+s^9+s^{10}=0, \quad (n=3) \\
 && 1-3s^5-18s^6+12s^7+12s^8+41s^9-66s^{10}+9s^{11}-9s^{12}+48s^{13}-36s^{14} +8s^{15}=0,\nonumber \\
&& \qquad  \quad (n=4) \\
 && 1-3s^6-30s^7+60s^9-30s^{10}+216s^{11}-329s^{12}-48s^{13} +3s^{14}+396s^{15}  \nonumber \\
&& \qquad +180s^{16}
-1113s^{17}+1038s^{18}-393s^{19} +48s^{20}+3s^{21}=0,\quad  (n=5) \\
 &&1-3s^7-45s^8-45s^9+165s^{10}-75s^{11}+165s^{12}+510s^{13}-1056s^{14}-959s^{15}+367s^{16}\nonumber\\
&&\quad +2349s^{17}+3433s^{18}-6589s^{19}-9069s^{20}+22070s^{21}-11495s^{22}-3597s^{23}+4455s^{24}\nonumber\\
&& \quad +702s^{25}-1971s^{26}+792s^{27}-106s^{28}=0, \quad (n=6).
\eea
The threshold $s_c=1-2\sin (\pi/18)$ for the $n=1$ kagome lattice
was first  given in \cite{kagomesite}. 
The threshold $s_c=1/\sqrt 2$ for $n=2$ has also been obtained by a ``cell-to-cell" transformation in \cite{akagome21}. 
Here, the thresholds for $3\leq n \leq  6$ are new.

We also computed $s_c$ and other critical properties numerically. The  results are summarized in Table \ref{pcsite}. 
Again, our numerical estimates of $s_c$ agree with the exact results up to 7 decimal places, and
these models all belong to the two-dimensional $q=1$ Potts model universality class.

Finally, we comment on some numerical specifics. Since the  number of non-crossing partitions 
  for site percolation is 
much larger than that of the  Potts model in both the magnetic and non-magnetic sector
for a given circumference $L$, the maximum 
system size $L=12$ that we reached is smaller.
The largest dimension of arrays used to save values and  positions of nonzero elements of 
the sparse matrices is $d^{S (m)}_{L=13}=125481607$, which requires about 43 gigabytes computer memory in the calculations. 
Corrections to scaling due to the leading irrelevant field is about $-1.8(1)$ for all lattices.


%
%
%
%

\section{Summary}
\label{summary}
We have studied critical properties of the $q$-state Potts model and bond and site percolation on two general classes of
lattices, the triangular-type and kagome-type lattices. 
For the triangular-type lattices of Fig. 1(a), the exact critical
frontier is known and  this led to a determination of the exact critical  thresholds of  site
percolation on $(1\times 1):(n\times n)$ kagome-type subnet lattices. Results for 
$1\leq n \leq 6$ are given.  
 For the kagome-type lattices of Fig. 1(b), no exact results are known except for $q=2$.
We carried out 
finite-size analysis to numerically determine critical properties for various lattice models 
including the 3-12 lattice and  $(n\times n):(n\times n)$ kagome-type subnet lattices.
Our numerical results on  conformal anomaly and 
magnetic correlation length verify that the principle of universality holds.

We have also computed the critical thresholds  for the Potts and bond percolation on
the 3-12 lattice and   $(n\times n):(n\times n)$ kagome-type subnet
 lattices using the homogeneity assumption  (\ref{conj}). 
To assess the accuracy of our numerical analysis as well as that of the
homogeneity assumption, we have applied our numerical procedure
to study critical properties of models for which exact results are known. 
The comparison of numerical and known results shows that
the numerical procedure is  accurate to 7 or 8 significant digits in determining 
critical thresholds. 
Assuming the same degree of accuracy for all lattices, this in turn
infers that the homogeneity assumption determines critical threshold with an accuracy up to 
5 decimal places or higher. 

Finally, our  analysis of critical properties 
is based on the use of lattice-dependent constants $A, B, C$ for the hatched triangles 
shown in Fig. 1.
We have developed an   algorithm of evaluating expressions of 
 $A,B,C$ using computers for hatched triangles  in the form of a stack-of-triangle structure.

\section*{Acknowledgment}
FYW would like to thank Professor Z. G. Zheng for the hospitality at the Beijing 
Normal University.
WG is much indebted to H. W. J. Bl\"ote for valuable discussions.
We thank R. M. Ziff for valuable comments and sending a copy of Ref. \cite{SZ} prior to publication.
This work is supported by the National Science Foundation of China (NSFC) under Grant No. 10675021, 
by the Program for New Century Excellent Talents in University (NCET),
and by the High Performance Scientific Computing Center (HSCC) of the Beijing 
Normal University. 

\appendix*
\section{ Expressions of $A, B, C$}
In this Appendix we list  constants $A, B, C$ computed using the computer algorithm as described in Sec. III.
The condition (\ref{condition}) holds for  for ferromagnetic Potts models with $v, m, q >0$.
 This confirms that  $qA=C$ is the exact critical frontier in the ferromagnetic regime. 
\subsection{Potts model on  $n\times n$ subnets with pure 2-site coupling $K$} 
  $v=e^{K}-1$

{\bf Subnet $1\times 1$}:
\begin{eqnarray}
A=1, \ \ \ B=v, \ \ \ C=v^3+3v^2. \label{1times1}
\end{eqnarray}

{\bf Subnet $2\times 2$}:
\begin{eqnarray}
A&=&q^{3}+9q^{2}v+33qv^{2}+(50+4q)v^{3}+21v^{4}+3v^{5} \nonumber \\
B&=&q^{2}v^{2}+10qv^{3}+(30+2q)v^{4}+22v^{5}+7v^{6}+v^{7}\nonumber \\
C&=&9qv^{4}+(54+3q)v^{5}+63v^{6}+33v^{7}+9v^{8}+v^{9}.
\end{eqnarray}

{\bf Subnet $3\times 3$}:
\begin{eqnarray}
A&=&29q^{3}v^{6}+(459q^{2}+9q^{3})v^{7}+(2592q+423q^{2})v^{8}+(5292+4185q+171q^{2})v^{9}\nonumber\\
 &+&(12825+3258q+36q^{2})v^{10}+(15534+1539q+3q^{2})v^{11}+(12184+454q)v^{12}\nonumber\\
 &+&(6732+78q)v^{13}+(2688+6q)v^{14}+768v^{15}+150v^{16}+18v^{17}+v^{18}, \nonumber \\
B&=&q^{5}v^{3}+21q^{4}v^{4}+(199q^{3}+3q^{4})v^{5}+(1040q^{2}+85q^{3})v^{6}+(2979q+844q^{2}\nonumber\\
 &+&17q^{3})v^{7}+(3780+3834q+340q^{2}+2q^{3})v^{8}+(7182+2429q+86q^{2})v^{9}\nonumber\\
 &+&(6858+950q+13q^{2})v^{10}+(4250+233q+q^{2})v^{11}+(1846+33q)v^{12}\nonumber\\
 &+&(570+2q)v^{13}+121v^{14}+16v^{15}+v^{16},\nonumber \\
C&=&29q^{3}v^{6}+(459q^{2}+9q^{3})v^{7}+(2592q+423q^{2})v^{8}+(5292+4185q+171q^{2})v^{9}\nonumber\\
 &+&(12825+3258q+36q^{2})v^{10}+(15534+1539q+3q^{2})v^{11}+(12184+454q)v^{12}\nonumber\\
 &+&(6732+78q)v^{13}+(2688+6q)v^{14}+768v^{15}+150v^{16}+18v^{17}+v^{18}.
\end{eqnarray}
{\bf Subnet $4\times 4$}:
\begin{eqnarray}
A&=&q^{12}+30q^{11}v+435q^{10}v^{2}+(4044q^{9}+16q^{10})v^{3}+(26952q^{8}+450q^{9})v^{4}+(136323q^{7}\nonumber\\
&+&6057q^{8}+18q^{9})v^{5}+(539687q^{6}+51643q^{7}+606q^{8})v^{6}+(1696074q^{5}+310764q^{6}\nonumber\\
&+&9336q^{7}+27q^{8})v^{7}+(4229307q^{4}+1388580q^{5}+88062q^{6}+948q^{7})v^{8}+(8218405q^{3}\nonumber\\
&+&4700302q^{4}+568443q^{5}+15274q^{6}+43q^{7})v^{9}+(11888619q^{2}+11989938q^{3}\nonumber\\
&+&2635653q^{4}+148602q^{5}+1608q^{6})v^{10}+(11554638q+22208904q^{2}+8874255q^{3}\nonumber\\
&+&966189q^{4}+26310q^{5}+96q^{6})v^{11}+(5728860+27132098q+21124723q^{2}+4354609q^{3}\nonumber\\
&+&251967q^{4}+3222q^{5}+3q^{6})v^{12}+(16691130+32535501q+13440978q^{2}+1559913q^{3}\nonumber\\
&+&48486q^{4}+267q^{5})v^{13}+(24925347+26263347q+6337347q^{2}+426225q^{3}+6834q^{4}\nonumber\\
&+&12q^{5})v^{14}+(25218686+15860794q+2323668q^{2}+89613q^{3}+644q^{4})v^{15}\nonumber\\
&+&(19264962+7537923q+674634q^{2}+14088q^{3}+30q^{4})v^{16}+(11718378+2882733q\nonumber\\
&+&154248q^{2}+1539q^{3})v^{17}+(5825765+890008q+27036q^{2}+102q^{3})v^{18}+(2389554\nonumber\\
&+&219258q+3447q^{2}+3q^{3})v^{19}+(806778+42009q+288q^{2})v^{20}+(221570+5996q\nonumber\\
&+&12q^{2})v^{21}+(48435+594q)v^{22}+(8136+36q)v^{23}+(990+q)v^{24}+78v^{25}+3v^{26},\nonumber 
\end{eqnarray}
\begin{eqnarray}
B&=&q^{9}v^{4}+36q^{8}v^{5}+(609q^{7}+4q^{8})v^{6}+(6340q^{6}+193q^{7})v^{7}+(44883q^{5}+3730q^{6}\nonumber\\
&+&30q^{7})v^{8}+(225145q^{4}+40957q^{5}+1156q^{6}+3q^{7})v^{9}+(804174q^{3}+290085q^{4}\nonumber\\
&+&19521q^{5}+240q^{6})v^{10}+(1980543q^{2}+1384197q^{3}+192846q^{4}+6208q^{5}+30q^{6})v^{11}\nonumber\\
&+&(3064302q+4394962q^{2}+1220478q^{3}+85315q^{4}+1355q^{5}+2q^{6})v^{12}+(2280420\nonumber\\
&+&8560443q+4979114q^{2}+717391q^{3}+26906q^{4}+201q^{5})v^{13}+(7901226+12201635q\nonumber\\
&+&3771317q^{2}+307105q^{3}+6145q^{4}+19q^{5})v^{14}+(14006718+11666444q+2106366q^{2}\nonumber\\
&+&99167q^{3}+989q^{4}+q^{5})v^{15}+(16765996+8289847q+906964q^{2}+24293q^{3}\nonumber\\
&+&102q^{4})v^{16}+(15077600+4592110q+306788q^{2}+4431q^{3}+5q^{4})v^{17}+(10735261\nonumber\\
&+&2026213q+81784q^{2}+574q^{3})v^{18}+(6217796+715829q+17041q^{2}+48q^{3})v^{19}\nonumber\\
&+&(2966339+200961q+2720q^{2}+2q^{3})v^{20}+(1168690+43935q+319q^{2})v^{21}\nonumber\\
&+&(378274+7210q+25q^{2})v^{22}+(99306+833q+q^{2})v^{23}+(20684+60q)v^{24}\nonumber\\
&+&(3298+2q)v^{25}+379v^{26}+28v^{27}+v^{28},\nonumber 
\end{eqnarray}
\begin{eqnarray}
C&=&99q^{6}v^{8}+(2871q^{5}+29q^{6})v^{9}+(36285q^{4}+2250q^{5})v^{10}+(256785q^{3}+47886q^{4} \nonumber\\
 &+&765q^{5})v^{11}+(1078677q^{2}+489164q^{3}+30525q^{4}+135q^{5})v^{12}+(2567538q \nonumber\\
 &+&2747088q^{2}+463617q^{3}+11898q^{4}+9q^{5})v^{13}+(2723220+8371674q  \nonumber\\
 &+&3520566q^{2}+282843q^{3}+2919q^{4})v^{14}+(11070162+13828185q+2965938q^{2}  \nonumber\\
 &+&120355q^{3}+420q^{4})v^{15}+(22921893+15205050q+1805958q^{2}+36504q^{3}+27q^{4})v^{16}  \nonumber\\
 &+&(31898508+12327042q+826827q^{2}+7773q^{3})v^{17}+(33199952+7729828q+287940q^{2}  \nonumber\\
 &+&1098q^{3})v^{18}+(27253662+3833517q+75627q^{2}+90q^{3})v^{19}+(18157536+1513779q  \nonumber\\
 &+&14556q^{2}+3q^{3})v^{20}+(9965342+473460q+1938q^{2})v^{21}+(4531923+115287q  \nonumber\\
 &+&159q^{2})v^{22}+(1706052+21150q+6q^{2})v^{23}+(527795+2757q)v^{24}+(132300+228q)v^{25}  \nonumber\\
 &+&(26256+9q)v^{26}+3976v^{27}+432v^{28}+30v^{29}+v^{30}
\end{eqnarray}

\subsection{Potts model on $n\times n$ subnets with pure 3-site coupling $M$}
 $m=e^M-1$

 {\bf Subnet $1\times 1$}:
\begin{eqnarray}
A=1, \ \  \ B=0, \ \ \ C=m.
\end{eqnarray}

{\bf Subnet $2\times 2$}:
\begin{eqnarray}
A=  q^{3}+3qm,\ \ \ B= m^{2}, \ \ \ C= m^{3}.
\end{eqnarray}

{\bf Subnet $3\times 3$}:
\begin{eqnarray}
A&=&q^{7}+6q^{5}m+15q^{3}m^{2}+(14q+3q^{2})m^{3}+3m^{4},\nonumber \\
B&=&q^{2}m^{3}+(2+2q)m^{4}+m^{5},\nonumber \\
C&=&3m^{5}+m^{6}.
\end{eqnarray}

{\bf Subnet $4 \times 4$}:
\begin{eqnarray}
A&=&q^{12}+10q^{10}m+45q^{8}m^{2}+(114q^{6}+6q^{7})m^{3}+(165q^{4}+42q^{5})m^{4}\nonumber\\
&+&(117q^{2}+99q^{3}+9q^{4})m^{5}+(20+73q+33q^{2}+3q^{3})m^{6}\nonumber\\
&+&(15+13q)m^{7}+3m^{8},\nonumber \\
B&=&q^{5}m^{4}+(6q^{3}+3q^{4})m^{5}+(9q+15q^{2}+3q^{3})m^{6}+(12+12q+3q^{2})m^{7}\nonumber\\
&+&(6+2q)m^{8}+m^{9}\nonumber \\
C&=&(2+9q)m^{7}+(15+3q)m^{8}+7m^{9}+m^{10}.
\end{eqnarray}

{\bf Subnet $5 \times 5$}:
\begin{eqnarray}
A&=&q^{18}+15q^{16}m+105q^{14}m^{2}+(445q^{12}+10q^{13})m^{3}+(1245q^{10}+120q^{11})m^{4}+(2358q^{8}\nonumber \\
&+&624q^{9}+18q^{10})m^{5}+(2967q^{6}+1795q^{7}+189q^{8}+6q^{9})m^{6}+(2298q^{4}+2976q^{5}+792q^{6}\nonumber \\
&+&84q^{7})m^{7}+(888q^{2}+2613q^{3}+1608q^{4}+399q^{5}+27q^{6})m^{8}+(86+864q+1416q^{2}\nonumber \\
&+&808q^{3}+182q^{4}+9q^{5})m^{9}+(249+579q+345q^{2}+81q^{3}+3q^{4})m^{10}+(126+129q\nonumber \\
&+&36q^{2})m^{11}+(30+13q)m^{12}+3m^{13},\nonumber
\end{eqnarray}
\begin{eqnarray}
B&=&q^{9}m^{5}+(12q^{7}+4q^{8})m^{6}+(47q^{5}+42q^{6}+6q^{7})m^{7}+(80q^{3}+156q^{4}+57q^{5}+7q^{6})m^{8}\nonumber \\
&+&(42q+221q^{2}+204q^{3}+59q^{4}+7q^{5})m^{9}+(67+225q+169q^{2}+58q^{3}+5q^{4})m^{10}\nonumber \\
&+&(105+121q+37q^{2}+5q^{3})m^{11}+(49+25q+3q^{2})m^{12}+(11+2q)m^{13}+m^{14}, \nonumber
\end{eqnarray}
\begin{eqnarray}
C&=&(12q^{2}+29q^{3})m^{9}+(6+63q+96q^{2}+9q^{3})m^{10}+(102+132q+36q^{2})m^{11}+(127+51q\nonumber \\
&+&3q^{2})m^{12}+(57+6q)m^{13}+12m^{14}+m^{15}.
\end{eqnarray}

{\bf Subnet $6 \times 6$}:
\begin{eqnarray}
A&=&q^{25}+21q^{23}m+210q^{21}m^{2}+(1315q^{19}+15q^{20})m^{3}+(5715q^{17}+270q^{18})m^{4}\nonumber \\
&+&(18084q^{15}+2235q^{16}+30q^{17})m^{5}+(42560q^{13}+11166q^{14}+525q^{15}+10q^{16})m^{6}\nonumber \\
&+&(74769q^{11}+37113q^{12}+4113q^{13}+210q^{14})m^{7}+(96819q^{9}+85161q^{10}+18795q^{11}\nonumber \\
&+&1926q^{12}+54q^{13})m^{8}+(89350q^{7}+135082q^{8}+54561q^{9}+10086q^{10}+798q^{11}\nonumber \\
&+&18q^{12})m^{9}+(54795q^{5}+143112q^{6}+102321q^{7}+32634q^{8}+5172q^{9}+351q^{10}\nonumber \\
&+&6q^{11})m^{10}+(19257q^{3}+92445q^{4}+118923q^{5}+65685q^{6}+18570q^{7}+2622q^{8}\nonumber \\
&+&174q^{9})m^{11}+(2642q+29213q^{2}+74961q^{3}+76633q^{4}+38522q^{5}+9990q^{6}+1452q^{7}\nonumber \\
&+&66q^{8})m^{12}+(2106+17556q+41547q^{2}+42069q^{3}+20271q^{4}+5523q^{5}+735q^{6}\nonumber \\
&+&27q^{7})m^{13}+(4536+16626q+18789q^{2}+10158q^{3}+2817q^{4}+381q^{5}+9q^{6})m^{14}\nonumber \\
&+&(3459+6819q+4372q^{2}+1398q^{3}+179q^{4}+3q^{5})m^{15}+(1398+1593q+594q^{2}\nonumber \\
&+&87q^{3})m^{16}+(342+216q+36q^{2})m^{17}+(48+13q)m^{18}+3m^{19},\nonumber 
\end{eqnarray}
\begin{eqnarray}
B&=&q^{14}m^{6}+(20q^{12}+5q^{13})m^{7}+(145q^{10}+90q^{11}+10q^{12})m^{8}+(540q^{8}\nonumber \\
&+&623q^{9}+168q^{10}+14q^{11})m^{9}+(1142q^{6}+2263q^{7}+1132q^{8}+221q^{9}+17q^{10})m^{10}\nonumber \\
&+&(1304q^{4}+4549q^{5}+4089q^{6}+1413q^{7}+258q^{8}+16q^{9})m^{11}+(659q^{2}+4573q^{3}\nonumber \\
&+&7808q^{4}+4873q^{5}+1569q^{6}+238q^{7}+16q^{8})m^{12}+(68+1652q+6525q^{2}+8325q^{3}\nonumber \\
&+&4896q^{4}+1432q^{5}+214q^{6}+14q^{7})m^{13}+(1183+4981q+6786q^{2}+3999q^{3}+1184q^{4}\nonumber \\
&+&182q^{5}+11q^{6})m^{14}+(1950+4062q+2733q^{2}+885q^{3}+141q^{4}+9q^{5})m^{15}+(1372\nonumber \\
&+&1566q+571q^{2}+102q^{3}+7q^{4})m^{16}+(538+338q+64q^{2}+5q^{3})m^{17}+(127+40q\nonumber \\
&+&3q^{2})m^{18}+(17+2q)m^{19}+m^{20},\nonumber
\end{eqnarray}
\begin{eqnarray}
C&=&(54q^{5}+99q^{6})m^{11}+(146q^{3}+585q^{4}+483q^{5}+29q^{6})m^{12}+(90q+888q^{2}\nonumber\\
&+&1971q^{3}+1170q^{4}+159q^{5})m^{13}+(228+1992q+3303q^{2}+2073q^{3}+381q^{4}\nonumber\\
&+&9q^{5})m^{14}+(1677+3726q+2598q^{2}+660q^{3}+33q^{4})m^{15}+(2421+2511q\nonumber\\
&+&810q^{2}+81q^{3})m^{16}+(1626+813q+114q^{2}+3q^{3})m^{17}+(620+133q+6q^{2})m^{18}\nonumber\\
&+&(141+9q)m^{19}+18m^{20}+m^{21}.
\end{eqnarray}

{\bf Subnet $7\times 7$}:
\begin{eqnarray}
A&=&q^{33}+28q^{31}m+378q^{29}m^{2}+(3255q^{27}+21q^{28})m^{3}+(19950q^{25}\nonumber\\
&+&525q^{26})m^{4}+(92025q^{23}+6210q^{24}+45q^{25})m^{5}+(329615q^{21}+45955q^{22}\nonumber\\
&+&1155q^{23}+15q^{24})m^{6}+(932792q^{19}+237060q^{20}+13755q^{21}+430q^{22})m^{7}\nonumber\\
&+&(2102508q^{17}+899163q^{18}+100401q^{19}+5835q^{20}+90q^{21})m^{8}+(3777012q^{15}\nonumber\\
&+&2577591q^{16}+499731q^{17}+48899q^{18}+2062q^{19}+30q^{20})m^{9}+(5372928q^{13}\nonumber\\
&+&5650464q^{14}+1784592q^{15}+278639q^{16}+22401q^{17}+846q^{18}+10q^{19})m^{10}\nonumber\\
&+&(5963664q^{11}+9467118q^{12}+4680099q^{13}+1127826q^{14}+150330q^{15}+10692q^{16}\nonumber\\
&+&396q^{17})m^{11}+(5030270q^{9}+11966359q^{10}+9054966q^{11}+3301756q^{12}\nonumber\\
&+&683417q^{13}+80559q^{14}+5754q^{15}+132q^{16})m^{12}+(3081149q^{7}+11079301q^{8}\nonumber\\
&+&12758268q^{9}+6987391q^{10}+2179863q^{11}+399888q^{12}+46751q^{13}+2766q^{14}\nonumber\\
&+&54q^{15})m^{13}+(1263348q^{5}+7101903q^{6}+12629796q^{7}+10471011q^{8}\nonumber\\
&+&4893192q^{9}+1358682q^{10}+242670q^{11}+26103q^{12}+1422q^{13}+18q^{14})m^{14}\nonumber\\
&+&(294802q^{3}+2824768q^{4}+8154549q^{5}+10567810q^{6}+7525012q^{7}+3161463q^{8}\nonumber\\
&+&844184q^{9}+144491q^{10}+15006q^{11}+690q^{12}+6q^{13})m^{15}+(25948q+547184q^{2}\nonumber\\
&+&2935194q^{3}+6464096q^{4}+7397804q^{5}+4847082q^{6}+1962007q^{7}+511413q^{8}\nonumber\\
&+&86643q^{9}+8408q^{10}+348q^{11})m^{16}+(23688+392541q+1878534q^{2}+3986040q^{3}\nonumber\\
&+&4432011q^{4}+2878479q^{5}+1160562q^{6}+304893q^{7}+50895q^{8}+4761q^{9}\nonumber\\
&+&153q^{10})m^{17}+(112270+781871q+1899615q^{2}+2296835q^{3}+1561921q^{4}\nonumber\\
&+&654696q^{5}+175858q^{6}+29445q^{7}+2616q^{8}+66q^{9})m^{18}+(162018+662861q\nonumber\\
&+&987349q^{2}+764529q^{3}+345520q^{4}+97287q^{5}+16524q^{6}+1404q^{7}+27q^{8})m^{19}\nonumber\\
&+&(123912+327444q+320334q^{2}+166830q^{3}+50811q^{4}+8838q^{5}+747q^{6}\nonumber\\
&+&9q^{7})m^{20}+(60508+106552q+69966q^{2}+24288q^{3}+4547q^{4}+372q^{5}\nonumber\\
&+&3q^{6})m^{21}+(20268+23740q+10182q^{2}+2142q^{3}+188q^{4})m^{22}+(4710\nonumber\\
&+&3525q+894q^{2}+87q^{3})m^{23}+(732+316q+36q^{2})m^{24}+(69+13q)m^{25}+3m^{26},\nonumber
\end{eqnarray}
\begin{eqnarray}
B&=&q^{20}m^{7}+(30q^{18}+6q^{19})m^{8}+(345q^{16}+165q^{17}+15q^{18})m^{9}\nonumber\\
&+&(2175q^{14}+1820q^{15}+390q^{16}+25q^{17})m^{10}+(8587q^{12}+11240q^{13}+4199q^{14}\nonumber\\
&+&624q^{15}+35q^{16})m^{11}+(22421q^{10}+43891q^{11}+25772q^{12}+6546q^{13}+844q^{14}\nonumber\\
&+&40q^{15})m^{12}+(38966q^{8}+113011q^{9}+100896q^{10}+39578q^{11}+8607q^{12}+951q^{13}\nonumber\\
&+&43q^{14})m^{13}+(43431q^{6}+190110q^{7}+258498q^{8}+153642q^{9}+50665q^{10}+9597q^{11}\nonumber\\
&+&981q^{12}+44q^{13})m^{14}+(28116q^{4}+196731q^{5}+420318q^{6}+385688q^{7}+190950q^{8}\nonumber\\
&+&55686q^{9}+9626q^{10}+970q^{11}+40q^{12})m^{15}+(8320q^{2}+108004q^{3}+395182q^{4}\nonumber\\
&+&590541q^{5}+455390q^{6}+203444q^{7}+54523q^{8}+9175q^{9}+874q^{10}+37q^{11})m^{16}\nonumber\\
&+&(488+20976q+170827q^{2}+477931q^{3}+626376q^{4}+452018q^{5}+190124q^{6}+50024q^{7}\nonumber\\
&+&8064q^{8}+775q^{9}+32q^{10})m^{17}+(15646+137655q+396335q^{2}+531376q^{3}+381398q^{4}\nonumber\\
&+&162295q^{5}+42103q^{6}+6904q^{7}+640q^{8}+28q^{9})m^{18}+(51536+224439q+352647q^{2}\nonumber\\
&+&280816q^{3}+123719q^{4}+33731q^{5}+5505q^{6}+532q^{7}+22q^{8})m^{19}+(66881+176633q\nonumber\\
&+&170578q^{2}+86821q^{3}+24401q^{4}+4339q^{5}+407q^{6}+18q^{7})m^{20}+(49197+83330q\nonumber\\
&+&51390q^{2}+17058q^{3}+3016q^{4}+326q^{5}+13q^{6})m^{21}+(23535+25741q+10127q^{2}\nonumber\\
&+&2132q^{3}+217q^{4}+11q^{5})m^{22}+(7727+5333q+1281q^{2}+155q^{3}+7q^{4})m^{23}\nonumber\\
&+&(1751+719q+94q^{2}+5q^{3})m^{24}+(264+57q+3q^{2})m^{25}+(24+2q)m^{26}+m^{27},\nonumber
\end{eqnarray}
\begin{eqnarray}
C&=&(222q^{9}+351q^{10})m^{13}+(1386q^{7}+3816q^{8}+2250q^{9}+99q^{10})m^{14}\nonumber\\
&+&(3198q^{5}+14696q^{6}+18894q^{7}+7530q^{8}+686q^{9})m^{15}+(3072q^{3}+24933q^{4}\nonumber\\
&+&59031q^{5}+51366q^{6}+18519q^{7}+2208q^{8}+29q^{9})m^{16}+(888q+16710q^{2}+78501q^{3}\nonumber\\
&+&132330q^{4}+99747q^{5}+35370q^{6}+5022q^{7}+150q^{8})m^{17}+(1954+32712q+133203q^{2}\nonumber\\
&+&202318q^{3}+150051q^{4}+55086q^{5}+8997q^{6}+456q^{7})m^{18}+(25370+130809q+228960q^{2}\nonumber\\
&+&178833q^{3}+72045q^{4}+13566q^{5}+948q^{6}+9q^{7})m^{19}+(67692+179352q+174174q^{2}\nonumber\\
&+&78354q^{3}+17760q^{4}+1581q^{5}+33q^{6})m^{20}+(81987+127689q+73398q^{2}+19325q^{3}\nonumber\\
&+&2343q^{4}+72q^{5})m^{21}+(58543+54880q+18612q^{2}+2746q^{3}+150q^{4})m^{22}+(27453\nonumber\\
&+&15033q+2859q^{2}+207q^{3}+3q^{4})m^{23}+(8825+2607q+246q^{2}+6q^{3})m^{24}+(1949\nonumber\\
&+&264q+9q^{2})m^{25}+(285+12q)m^{26}+25m^{27}+m^{28}.
\end{eqnarray}

\end{document}